\documentclass[twocolumn,showpacs,eqsecnum,amsmath,amssymb,prb]{revtex4}

\usepackage{graphicx}
\usepackage{dcolumn}
\usepackage{bm}
\usepackage{psfrag}

\newcommand{\intd}{{\rm d}}
\newcommand{\imag}{{\rm i}}

\begin{document}

\preprint{}

\title{
Majorana zero modes
bound to a vortex line in a topological superconductor
}

\author{Takahiro  Fukui}
\affiliation{Department of Physics, Ibaraki University, Mito
310-8512, Japan}

\date{\today}

\begin{abstract}
We explore Majorana zero modes bound to a vortex line in 
a three dimensional topological superconductor model,
focusing our attention on the validity of the index theorem previously derived.
We first solve the Bogoliubov-de Gennes equation at the zero energy
to obtain the analytical index.
We next calculate the topological index given by the order parameters.
It turns out that they indeed coincide and that index theorem, which has been derived
on the implicit assumption that a defect is point-like,  
is also valid for a line defect.
\end{abstract}

\maketitle

\section{Introduction} 
There has been renewed interest recently in topological phases of matter, owing to the 
rapid progress in both theoretical and experimental studies of topological insulators.
\cite{HasKan10}
While the quantum Hall effect is a typical example of such a phase, \cite{TKNN82,Koh85}
recent discovery is that even with unbroken time reversal symmetry, 
band insulators can be topologically nontrivial. \cite{KanMel05a,KanMel05b,BerZha06,Roy09}
By extending the idea of the topological insulators to superconductors,
\cite{SRFL08,Kit08} 
it has been shown that various topological phases 
are classified by ten universality classes, similarly to 
the random matrix theory.\cite{Zir96,AltZir97}

Superconducting states are described by Bogoliubov-de Gennes Hamiltonians with
particle-hole symmetry. 
This symmetry guarantees a pair of states with energy $\pm E$ for every eigenstate.
The zero energy states then have a special property which enables us to regard them as
Majorana fermion modes. With many Majorana zero modes, 
the zero energy ground states are degenerate and 
these fermion modes are expected to obey the non-Abelian statistics.
They are predicted in many other systems such as a $p$-wave
superconductor, etc. 
\cite{ReaGre00,Iva01,SOM04,SNT06,TSL07,GurRad07,FuKan08,QHRZ09,STF09,TYN09,Vol09,BerHur09,SLTS10,Lee09,Ali10,Her10,LTYSN10,Her09b}
Generic conditions of the existence of the Majorana zero modes have been discussed.
\cite{Kit00,Kit06,TSS10,TeoKan10,HCM10,SNCM10,FukFuj10,SRCM09,SLTS09,Roy10} 
In particular, an index theorem has been derived,\cite{FukFuj10,Vol03} 
which reveals the relationship between 
the index of the Majorana zero modes (analytical index)
and a topological invariant given by the configuration 
of the order parameters (topological index) . 

Although previous studies have been focused on the non-Abelian statistics in lower 
spatial dimensions, Teo and Kane \cite{TeoKan10} have recently proposed
a three dimensional (3D) topological superconductor with a Majorana zero mode
and discussed the possibility of the non-Abelian statistics in three dimensions.
The index theorem has also been derived for the same system. \cite{FukFuj10}
Since a point defect has been considered implicitly in these works, 
it may be interesting to consider a line defect inherent in three dimensions and
to study whether Majorana modes appear or not in the core of a vortex line.
It is also nontrivial whether the 3D topological index
which is generically the winding number of the order parameter
on the two dimensional sphere 
gives a correct answer for a line defect which is expected to be classified by 
the winding number around a circle.

In this paper, we study a 3D topological superconductor model with a vortex line,
focusing our attention on the validity of the index theorem derived previously when
it is applied to the case of a line defect. 
In the next section, we discuss the model and its symmetry properties.
In Sec. \ref{s:MajZerMod}, 
we solve the Bogoliubov-de Gennes equation analytically to
show that in the core of the vortex line, there appear
Majorana bound states exactly at the zero energy.
We can then obtain the analytical index by counting the number of zero modes 
and their chirality. 
In Sec. \ref{s:IndThe},
we calculate the topological index in three dimensions in the case of a line defect.
We show that 
both the indices coincide indeed. 
Sec. \ref{s:Sum} is devoted to summary and discussions.

\section{A model and its symmetry}

We study in this paper a minimum model of a 3D topological superconductor 
proposed by Teo and Kane.\cite{TeoKan10} 
This model includes a mass controlling the band inversion, which  
is assumed to change its sign along the $z$-axis.
When the system is in the normal phase,
an ordinary insulating state appears in the positive mass domain, 
whereas a topological insulating state
is realized in the negative mass domain. At the boundary of these domains,  
massless Dirac fermions without doubling can appear.
As shown by Teo and Kane, when such a system is in the superconducting phase,
a topological defect  
such as a 't Hooft-Polyakov monopole
background field yields a Majorana zero mode.  
In what follows, we consider not a point defect but a line defect, i.e., a vortex line, and
show that the Majorana zero modes also appear in this model.

To be concrete, we study the following Hamiltonian,
\begin{alignat}1
&
{\cal H}={\cal H}_0+{\cal H}_1,
\nonumber\\
&{\cal H}_0=\imag\gamma^j\partial_j+\gamma^{j+3}\phi_j,
\nonumber\\
&
{\cal H}_1=\imag\mu\gamma^4\gamma^5 ,
\label{Ham}
\end{alignat}
where $j=1,2,3$, and the $\gamma$ matrices obey $\{\gamma^i,\gamma^j\}=2\delta^{ij}$
for $i,j=1,2,\cdots,6$.
We also define $\Gamma_5=(-\imag)^3\gamma^1\gamma^2\cdots\gamma^6$.
To solve the Bogoliubov-de Gennes equation for zero modes, 
it may be convenient to choose the $\gamma$ matrices such that
\begin{alignat}1
&\gamma^j=\sigma^1\otimes\sigma^j\otimes1,
\nonumber\\
&\gamma^{j+3}=\sigma^2\otimes1\otimes\sigma^j,
\nonumber\\
&\Gamma_5=\sigma^3\otimes1\otimes1,
\label{GamMat}
\end{alignat}
which is different from but unitary-equivalent to 
the representation used in Ref. \cite{TeoKan10}.
The term ${\cal H}_1$ in (\ref{Ham}) denotes the chemical potential term.

To describe a vortex line, we introduce the cylindrical coordinates $(r,\theta,z)$. 
Let $\phi_1$ and $\phi_2$ be the set of the superconducting order parameter with a vortex
which is a continuous function of $r$, $\theta$, and $z$. 
We assume its asymptotic behavior as
\begin{alignat}1
(\phi_1,\phi_2)\rightarrow\left\{
\begin{array}{ll}
\Delta_0\left(\cos q\theta,\sin q\theta\right)&(r\rightarrow\infty)\\
\Delta(r)\left(\cos q\theta,\sin q\theta\right)\quad&(|z|\rightarrow\infty)
\end{array}
\right. ,
\label{SCOrdParAsy}
\end{alignat}
where the consistency at $r\rightarrow\infty$ and $|z|\rightarrow\infty$ requires
that $\Delta(\infty)=\Delta_0$. Without loss of generality, we assume that $\Delta_0>0$.
Eq. (\ref{SCOrdParAsy}) tells that at $r\rightarrow\infty$, 
$(\phi_1,\phi_2)$ defines a map from S$^1$ to S$^1$.
Therefore, 
for the function $(\phi_1,\phi_2)$ to be continuous, 
some zero points, $(\phi_1,\phi_2)=(0,0)$, are indispensable
somewhere in the $(x,y)$ plane for every $z$. 
In particular, in the limit $|z|\rightarrow\infty$, we assume $\Delta(0)=0$.
Therefore, 
these zeros form lines connecting $r=0$ at $z\rightarrow\pm\infty$. 
As we shall show below, 
the details of these lines, for example, the shape of the lines and/or the number of
the lines, are irrelevant to the index.
A mass controlling the band inversion is, on the other hand, defined by
\begin{alignat}1
\phi_3=\phi_3(z) .
\label{BanInvOrdPar}
\end{alignat}
The mass is assumed asymptotically to be
\begin{alignat}1
\phi_3(\pm\infty)=\pm m_0 .
\label{AsyMas}
\end{alignat}
As we shall see later, only the sign of $m_0$ is relevant to the index. 
In passing, we would like to mention that we have neglected, for simplicity, 
the vector potential of an external magnetic field which produces a vortex.

Next, let us examine the symmetry of the model.
Define the particle-hole conjugation operator
\begin{alignat}1
{\cal C}=\imag\gamma^2\gamma^5 K, \quad {\cal C}^2=1.
\label{ParHolCon}
\end{alignat}
Then, for the Hamiltonian (\ref{Ham}), we see
\begin{alignat}1
{\cal C}{\cal H}{\cal C}^{-1}=-{\cal H},
\label{ParHolSym}
\end{alignat}
which implies that the model belongs to the class D, a most generic class
of superconductors.\cite{Zir96,AltZir97,SRFL08,Kit08} 
If the chemical potential ${\cal H}_1$ is neglected, the Hamiltonian
${\cal H}_0$ has higher symmetry. To see this, note
\begin{alignat}1
\Gamma_5{\cal H}_0\Gamma_5=-{\cal H}_0.
\label{ChiSym}
\end{alignat}
This tells chiral symmetry of ${\cal H}_0$, and the nonzero chemical potential
term ${\cal H}_1$ breaks it. These twofold symmetries of ${\cal H}_0$ lead to the following 
fictitious time reversal operator,
\begin{alignat}1
{\cal T}=\Gamma_5{\cal C}.
\nonumber
\end{alignat}
From the transformation properties (\ref{ParHolSym}) and (\ref{ChiSym}), as well as 
$\Gamma_5{\cal C}={\cal C}\Gamma_5$, it is readily seen that 
\begin{alignat}1
&
{\cal T}{\cal H}_0{\cal T}^{-1}={\cal H}_0, \quad {\cal T}^2=1.
\nonumber
\end{alignat}
Therefore, the model without the chemical potential term 
${\cal H}_0$ belongs to class BDI.\cite{PCFurusaki}

Although one may be interested in the Majorana zero modes mainly in the generic class of 
superconductors,
i.e., in class D, the additional chiral symmetry (\ref{ChiSym}) plays a crucial role 
in the discussion on the index theorem. 
Indeed, one can obtain zero modes even in the case of nonzero chemical potential,
but without chiral symmetry, it is not possible to define the index any longer.
Therefore, in the next section, we shall solve the zero modes separately for 
${\cal H}_0$ and ${\cal H}$.

\section{Majorana zero modes}\label{s:MajZerMod}

In order to solve the Bogoliubov-de Gennes equation, we specify  in this section 
the superconducting order parameter by extrapolating the boundary value (\ref{SCOrdParAsy})
to $r=0$ and/or $z=0$ simply as,
\begin{alignat}1
(\phi_1,\phi_2)=
\Delta(r)\left(\cos q\theta,\sin q\theta\right) .
\label{SCOrdPar}
\end{alignat}
Namely, we assume that Eq. (\ref{SCOrdPar}) is valid for any $r$, $\theta$, and $z$. 
We also assume that $\Delta(\infty)=\Delta_0$, and that $\Delta(0)=0$, 
implying that the core of a vortex is located just on the $z$ axis.
In this case, the Hamiltonian is decoupled into two parts, one depending on $r$ and $\theta$,
and the other depending on $z$ only.  
Such a model can be regarded as a quasi 2D model 
interpolating the 2D model proposed by Jackiw and Rossi \cite{JacRos81}
and 3D model by Jackiw and Rebbi.\cite{JacReb76}

We also note that the order parameters (\ref{SCOrdPar})
yields a continuous dynamical symmetry, which simplifies the calculations below.
Namely, the order parameters Eqs. (\ref{SCOrdPar}) and (\ref{BanInvOrdPar}),
and thus the Hamiltonian ${\cal H}$ are invariant under the rotation around the $z$ axis.
Indeed, we see 
\begin{alignat}1
[{\cal H}, {\cal J}_3]=0,
\label{J3ComRel}
\end{alignat}
where the angular momentum around the $z$ axis ${\cal J}_3$ is defined by
\begin{alignat}1
{\cal J}_3
&=-\imag\partial_\theta+\frac{1}{2}
\left(1\otimes\sigma^3\otimes1+q1\otimes1\otimes\sigma^3\right) 
\nonumber\\
&=-\imag\partial_\theta-\frac{\imag}{2}
\left(\gamma^1\gamma^2+q\gamma^4\gamma^5\right) .
\label{AngMomFul}
\end{alignat}
Note also that $[\Gamma_5,{\cal J}_3]=0$. It follows that the zero modes can be labeled by 
the chirality and the angular momentum.

In what follows, let us rewrite the Hamiltonian in a more convenient way to obtain the 
zero mode wave functions.
The representation of the $\gamma$ matrices in Eq. (\ref{GamMat}) enables us to
write the Hamiltonian (\ref{Ham}) as
\begin{alignat}1
&
{\cal H}=
\left(
\begin{array}{cc}-\mu1\otimes\sigma^3&\imag D^-\\\imag D^+&-\mu1\otimes\sigma^3\end{array}
\right),
\nonumber\\
&D^{\pm}=\sigma^j\otimes1\partial_j\pm 1\otimes\sigma^j\phi_j .
\nonumber
\end{alignat}
Correspondingly, the particle-hole conjugation operator (\ref{ParHolCon}) can be denoted as 
\begin{alignat}1
&
{\cal C}=
\left(
\begin{array}{cc}C&\\ & -C\end{array}
\right),
\nonumber\\
&C\equiv i\sigma^2\otimes i\sigma^2K,
\label{RedPHCon}
\end{alignat}
by means of which particle-hole symmetry (\ref{ParHolSym}) can be described such that
\begin{alignat}1
&
CD^\pm C^{-1}=-D^\pm.
\nonumber\\
&
C1\otimes\sigma^3C^{-1}=-1\otimes\sigma^3.
\nonumber
\end{alignat} 
The angular momentum ${\cal J}_3$ in Eq. (\ref{AngMomFul}) is also decomposed into
\begin{alignat}1
{\cal J}_3=1\otimes J_3,
\nonumber
\end{alignat}
where, $J_3$ is the block diagonal element of the angular momentum operator defined by
\begin{alignat}1
&
J_3=-\imag\partial_\theta +S_3,
\nonumber\\
&S_3=\frac{1}{2}\left(\sigma^3\otimes1+q1\otimes\sigma^3\right) .
\nonumber
\end{alignat}
$S_3$ is the sum of the real spin and the pseudo-spin due to the particle-hole space 
of the Nambu notation.
The commutation relation (\ref{J3ComRel}) is translated into
\begin{alignat}1
[D^\pm,J_3]=[\mu1\otimes\sigma^3,J_3]=0.
\label{J3ComRelRed}
\end{alignat}
Now define
\begin{alignat}1
&
U_m(\theta)\equiv e^{\imag(m-S_3)\theta},
\label{Um}
\end{alignat}
then, we see
\begin{alignat}1
J_3U_m(\theta)=mU_m(\theta).
\nonumber
\end{alignat}
This matrix shall be used to construct the eigen functions of the zero modes. Therefore, 
the single-valuedness of $U_m(\theta)$ under the $2\pi$ rotation with respect to $\theta$ 
requires that $m$ should take $m=(q+1)/2$ mod $1$.
This implies that a unit vortex with $q=1$ serves as a spin-1/2 object, and the total spin
is just the sum of the spin and the pseudo-spin. 
The stability of the Majorana modes against the finite chemical potential
depends on whether $m$ is an integer or half-integer, 
as we shall see momentarily.
It should also be noted that $C$ and $J_3$ anticommute:
$CJ_3=-J_3C$. 
It follows that
\begin{alignat}1
CU_mC^{-1}=U_{-m} .
\label{PHTraUm}
\end{alignat}
Therefore, 
the zero modes appear as a pair of states with angular momentum $\pm m$ except for
$m=0$. This exception plays a crucial role in the stability of the Majorana zero modes 
mentioned above.

\subsection{The case of $\mu=0$}

Since the model has chiral symmetry in the case of $\mu=0$, 
the Bogoliubov-de Gennes equation at the zero energy
is decoupled into two sectors with definite chiralities.
Set the wave function as $\Psi_m=(\Psi_m^+,\Psi_m^-)^T$,
where $\pm$ denote the chirality of the zero mode wave functions.
Then, the equation for the zero mode can be written as
\begin{alignat}1
D^\pm\Psi_m^\pm=0 .
\nonumber
\end{alignat}
Because of Eq. (\ref{J3ComRelRed}), 
the zero modes can be simultaneous eigenstates of the chirality 
and the angular momentum $J_3$. 
Therefore, we have introduced the quantum number $m$
which labels the angular momentum,
$J_3\Psi_m^\pm=m\Psi_m^\pm$.
It is easy to see that the unitary transformation $U_m(\theta)$ defined in
Eq. (\ref{Um}) removes the $\theta$-dependence from $D^\pm$:
\begin{widetext}
\begin{alignat}1
D_m^\pm &\equiv U_m^{-1}(\theta) D^\pm U_m(\theta)
\nonumber\\
&=
\left(
\begin{array}{cccc}
\partial_3\pm \phi_3&\pm\Delta&\partial_r+\frac{2m+(1-q)}{2r}&0\\
\pm\Delta&\partial_3\mp \phi_3&0&\partial_r+\frac{2m+(1+q)}{2r}\\
\partial_r-\frac{2m-(1+q)}{2r}&0&-(\partial_3\mp \phi_3)&\pm\Delta\\
0&\partial_r-\frac{2m-(1-q)}{2r}&\pm\Delta&-(\partial_3\pm \phi_3)
\end{array}
\right) .
\nonumber
\end{alignat}
\end{widetext}
Therefore, the transformed operator $D_m^\pm$ can be regarded as the operator
with a definite angular momentum $m$.
Correspondingly, the eigenstate of $J_3$ is defined as
\begin{alignat}1
&
\Psi_m^\pm(r,\theta,z)=U_m(\theta)\psi^\pm_m(r,z).
\nonumber 
\end{alignat}
Thus, it turns out that
the $\theta$-dependence of the zero mode wave functions is determined by $U_m$.

Next, note that the $z$-dependence is only from the 
diagonal elements of the Hamiltonian. To solve it, we set
\begin{alignat}1
&
\psi^+_m=(\eta_{1m}^+,\xi_{1m}^+,\xi_{2m}^+,\eta_{2m}^+)^T,
\nonumber\\
&
\psi^-_m=(\xi_{1m}^-,\eta_{1m}^-,\eta_{2m}^-,\xi_{2m}^-)^T.
\nonumber
\end{alignat}
Then, we have
\begin{alignat}1
&
\eta_m^\pm(r,z)=e^{-\int^z \intd z\phi_3(z)}\eta_m^\pm(r) ,
\nonumber\\
&
\xi_m^\pm(r,z)=e^{+\int^z \intd z\phi_3(z)}\xi_m^\pm(r) ,
\nonumber
\end{alignat}
where we have used the notation $\eta_m^\pm(r)=(\eta_{1m}^\pm(r),\eta_{2m}^\pm(r))^T$ and
$\xi_m^\pm(r)=(\xi_{1m}^\pm(r),\xi_{2m}^\pm(r))^T$.
The asymptotic mass $m_0$ in Eq. (\ref{AsyMas}) tells that the above 
two types of the wave functions,
$\eta$ and $\xi$, are normalizable only for $m_0>0$ and $m_0<0$, respectively.
Thus, the normalizability of the wave functions in the limit $|z|\rightarrow\infty$ requires
that the zero mode equation should reduce to
\begin{alignat}1
&
\xi^\pm_m(r)=0,\quad d_{m,\pm q}^\pm\eta^\pm_m(r)=0,\quad (m_0>0),
\nonumber\\
&
\eta^\pm_m(r)=0,\quad d_{m,\mp q}^\pm\xi^\pm_m(r)=0,\quad (m_0<0) ,
\label{RedZerModEqu}
\end{alignat}
where
\begin{alignat}1
d_{m,q}^\pm=
\left(
\begin{array}{cc}
\pm\Delta&\partial_r+\frac{2m+(1+q)}{2r} \\
\partial_r-\frac{2m-(1+q)}{2r} &\pm\Delta
\end{array}
\right) .
\nonumber
\end{alignat}

Finally, we examine the normalizability in the limit $r\rightarrow0$ and $\infty$. 
Let us consider the case of $m_0>0$, for example.
In the vicinity of $r=0$, we can set $\Delta(r)\sim0$, which implies that 
$\eta_{1m}^\pm$ and $\eta_{2m}^\pm$ are decoupled in Eq. (\ref{RedZerModEqu}). 
Then, we see that the wave function behaves such that 
$\eta_{1m}^\pm\sim r^{m-(1\pm q)/2}$ and
$\eta_{2m}^\pm\sim r^{-m-(1\pm q)/2}$.
For the wave function to be normalizable, we should impose the condition that
$r|\eta_{jm}^\pm|^2\propto r^\alpha$ with $\alpha>-1$ for both $j=1,2$.
We then see that
$\eta_m^\pm$ are normalizable, respectively, only for 
$|m|<(1\mp q)/2$.
Therefore, in the case of  $q>0$, we should impose $\eta_m^+=0$,
since $\eta_m^-$ is normalizable but $\eta_m^+$ is unnormalizable.
Likewise, in the case of $q<0$, we should impose $\eta_m^-=0$.

\begin{table}[ht]
\begin{ruledtabular}
\begin{tabular}{cccr}
$m_0$& $q$&  Nonzero wave fn
\footnote{The angular momentum $m$ takes $m=\frac{q+1}{2}$ mod 1 
in  $-\frac{|q|-1}{2}\le m \le \frac{|q|-1}{2}$. 
Therefore, in each case, there are $|q|$ normalizable zero modes.}& 
Index \footnote{The index of the Hamiltonian is defined in Eq. (\ref{DefInd}) }\\
\hline
$+$&$+$&$\eta^-_m$ &$-q$\\
$-$&$+$&$\xi^+_m$&$+q$\\
$+$&$-$&$\eta^+_m$&$|q|$\\
$-$&$-$&$\xi^-_m$&$-|q|$\\
\end{tabular}
\caption{Nonzero components of the normalizable wave functions}
\label{t:NorWavFun}
\end{ruledtabular}
\end{table}

It should be noted that the wave function discussed above 
has still an ambiguity with respect to the ratio of the two components 
$\eta_{1m}^\pm$ and $\eta_{2m}^\pm$ for each allowed $m$. 
This parameter plays a role in the normalizability in the limit, $r\rightarrow\infty$.
In this limit, $1/r$ terms in Eq. (\ref{RedZerModEqu}) can be neglected, and it is easy to show 
that the solution of the equation is a linear combination of $e^{\pm\Delta_0r}$.
Therefore, one can fine-tune the remaining parameter to vanish the exponentially 
growing component $e^{+\Delta_0 r}$.
This is the normalizable wave function at $r\rightarrow\infty$.

Final results, including the case of $m_0<0$, 
are summarized in Table \ref{t:NorWavFun}. 

\subsection{The case of $\mu\ne0$}
Since chiral symmetry is broken by the nonzero chemical potential,
the wave functions $\psi^\pm_m(r,z)$ couple together when $\mu\ne0$.
Even in this case, however,  
$\eta$ and $\xi$ do not coexist in the same wave function,
owing to the normalizability in the limit $|z|\rightarrow \infty$.
The zero mode equation is thus given by
\begin{alignat}1
\xi^\pm_m=0,\quad
&
\left(
\begin{array}{cc}
i\mu\sigma^3& d^-_{m,-q}\\d^+_{m,+q}&-i\mu\sigma^3
\end{array}
\right)
\left(
\begin{array}{c}\eta_m^+\\\eta_m^-\end{array}
\right)=0,\quad (m_0>0),
\nonumber\\
\eta^\pm_m=0,\quad
&
\left(
\begin{array}{cc}
-i\mu\sigma^3& d^-_{m,+q}\\d^+_{m,-q}&i\mu\sigma^3
\end{array}
\right)
\left(
\begin{array}{c}\xi_m^+\\\xi_m^-\end{array}
\right)=0,\quad (m_0<0).
\label{ZerModEquMu}
\end{alignat}
These equations do not allow normalizable solutions in general.
However, one exception is the case $m=0$, which occurs only when $q=$ odd.
To see this, note that Eqs. (\ref{RedPHCon}) and (\ref{PHTraUm}) enable us to impose
the relationship 
$\pm C\psi_m^{\pm}(r,z)=\psi_{-m}^\pm(r,z)$.
Hence, $m=0$ is quite special in the sense that independent variables (the components of 
$\eta_0^\pm$ or $\xi_0^\pm$) are reduced:
$\eta_{20}^\pm=\eta_{10}^{\pm*}$ and $\xi_{20}^\pm=-\xi_{10}^{\pm*}$.
Therefore, we have for $m_0>0$
\begin{alignat}1
&
\left(\partial_r+\frac{1+q}{2r}\right)\eta_{10}^++\Delta\eta_{10}^{+*}+i\mu\eta_{10}^{-*}=0,
\nonumber\\
&
\left(\partial_r+\frac{1-q}{2r}\right)\eta_{10}^{-}-\Delta\eta_{10}^{-*}-i\mu\eta_{10}^{+*}=0,
\nonumber
\end{alignat}
and for $m_0<0$
\begin{alignat}1
&
\left(\partial_r+\frac{1+q}{2r}\right)\xi_{10}^{-}+\Delta\xi_{10}^{-*}-i\mu\xi_{10}^{+*}=0,
\nonumber\\
&
\left(\partial_r+\frac{1-q}{2r}\right)\xi_{10}^{+}-\Delta\xi_{10}^{+*}+i\mu\xi_{10}^{-*}=0.
\nonumber
\end{alignat}
The normalizable solution is for $m_0>0$
\begin{alignat}1
\left(
\begin{array}{l}
\eta_{10}^+(r)\\ \eta_{10}^-(r)
\end{array}
\right)
\propto
e^{-\int^r\intd r\Delta(r)}
\left(
\begin{array}{r}
J_{(q+1)/2}(\mu r)\\-{\rm i}J_{(q-1)/2}(\mu r)
\end{array}
\right) ,
\nonumber
\end{alignat}
and for $m_0<0$
\begin{alignat}1
\left(
\begin{array}{l}
\xi_{10}^-(r)\\ \xi_{10}^+(r)
\end{array}
\right)
\propto
e^{-\int^r\intd r\Delta(r)}
\left(
\begin{array}{r}
J_{(q+1)/2}(\mu r)\\{\rm i}J_{(q-1)/2}(\mu r)
\end{array}
\right) ,
\nonumber
\end{alignat}
where $J_\nu(x)$ is the Bessel function of the first kind.
This result implies that if one adds chiral symmetry breaking perturbations,
only one zero mode survives when $q$ is odd, whereas no zero modes survive when 
$q$ is even. In this sense, the Majorana zero mode in class D is classified by Z$_2$.
\cite{TSL07,GurRad07,CLGS09,FukFuj10}

In passing, we mention that the zero mode solution obtained in this subsection is
nothing to do with the index, since it cannot be an eigen state of the chirality
$\Gamma_5$.

\section{Index theorem}\label{s:IndThe}

The index is associated with the zero modes with definite chiralities,
reflecting an analytical property of the Hamiltonian as a differential operator,
and the index theorem relates the index with a topological invariant given by
the order parameter.
In this section, we first review the index theorem obtained previously, \cite{FukFuj10}
and next show that the analytical index and
the topological index of the present model indeed coincide exactly.

It should be noted that chiral symmetry plays a central role in the index theorem.
Therefore, we restrict our discussions in this section to the Hamiltonian ${\cal H}_0$.

\subsection{Summary of index theorem}

The index of ${\cal H}_0$ is defined as 
\begin{alignat}1
{\rm ind}\,{\cal H}_0=N_+-N_-,
\label{DefInd}
\end{alignat}
where $N_\pm$ is respectively the number of zero modes with chirality $\pm$.
The chirality is here defined as the eigenvalue of $\Gamma_5$.
From the zero modes obtained in Sec. \ref{s:MajZerMod}, we can compute the index, which 
is summarized in Table \ref{t:NorWavFun}.
It has been shown that the index of ${\cal H}_0$ can be expressed by
\cite{Wei81,FukFuj10} 
\begin{alignat}1
{\rm ind}\,{\cal H}_0=-\frac{1}{2}\int \intd S_j J^j(x,0,\infty),
\label{IndThe}
\end{alignat}
where $\intd S_j$ is the infinitesimal surface element 
at the boundary of R$^d$ (typically S$^{d-1}$ at $r\rightarrow\infty$),
and $J^j(x,M_0,M_\infty)$ is the axial-vector current
defined by
\begin{alignat}1
&
J^j(x,M_0,M_\infty)
\nonumber\\
&
=\lim_{y\rightarrow x}{\rm tr}\,\gamma_5\gamma^j
\left(
\frac{1}{M_0-\imag{\cal H}}-\frac{1}{M_\infty-\imag{\cal H}}
\right)\delta(x-y) .
\nonumber
\end{alignat}
Since the current itself is not well-defined, a Pauli-Villars regulator
with a mass $M_\infty$ has been taken into account.  
In three dimensions, we have, \cite{Cal78,FukFuj10}
in the limits $M_0\rightarrow0$ and $M_\infty\rightarrow\infty$ 
\begin{alignat}1
J^i(x,0,\infty)=\frac{1}{4\pi \phi^3}
\epsilon^{ijk}\epsilon^{abc}
\phi_{a}\partial_{j}\phi_{b}\partial_{k}\phi_{c},
\nonumber
\end{alignat}
where $i,..,a,..=1,2,3$ and
$\phi^2=\sum_j\phi_j^2$. In Ref. \cite{FukFuj10}, we assumed that
$\phi^2$ is constant at the boundary S$^2$ of R$^3$ 
(in the limit $r=\sqrt{x^2+y^2+z^2}\rightarrow\infty$), 
which implies that the singularity is due to a point defect and 
the system is spherically symmetric at infinity.
In the present case, however, 
the vortex is a line defect in the superconducting order parameter
$(\phi_1,\phi_2)$, 
and the system has cylindrical symmetry rather than the spherical symmetry.
Therefore, $\phi^2$ cannot be uniform at the boundary
$r=\sqrt{x^2+y^2}\rightarrow\infty$ or $|z|\rightarrow\infty$.
The index theorem (\ref{IndThe}) is nevertheless valid if the integral 
is performed over an appropriate surface,
a cylinder for the present model, reflecting the symmetry of the model.

\subsection{Calculation of topological index}

It is noted that the topological index, the right hand side of Eq. (\ref{IndThe}), 
is determined only by the boundary value of the order parameter.
Therefore, 
in the following discussions, the superconducting order parameter $(\phi_1,\phi_2)$ 
is not necessarily given by Eq. (\ref{SCOrdPar}): Only the assumption of 
Eq. (\ref{SCOrdParAsy}) is enough to compute the index, and therefore, even if the 
a vortex line is curved or deformed, 
the topological index is the same, and so should be the analytical index,
provided that the order parameter obeys Eq. (\ref{SCOrdParAsy}) at the boundary.
This is one of advantages to utilize the topological index rather that the analytic index. 

In order to calculate the integral over a cylindrical surface in Eq. (\ref{IndThe}),
it should be noted that in terms of the form, the surface element $\intd S_j$ is 
given by the following two-form, $\intd S_j=(1/2!)\epsilon_{jkl}\intd x^k\intd x^l$,
where ${\rm d}x^k{\rm d}x^l=-{\rm d}x^l{\rm d}x^k$, and
that the order parameter depends on the cylindrical coordinates such that 
$\bm \phi=(\phi_1(r,\theta),\phi_2(r,\theta),\phi_3(z))$ at the boundary.
On the $(\theta,z)$ surface in the limit $r\rightarrow\infty$,
we have $J^j\intd S_j=(J^1\intd x^2-J^2\intd x^1)\intd x^3$.
Here, the current at the boundary $r\rightarrow\infty$ is given by 
\begin{alignat}1
J^{i}=\frac{1}{2\pi}\frac{\phi_3'(z)}{(\Delta_0^2+\phi_3^2(z))^{3/2}}
\epsilon^{ij}\epsilon^{ab}\phi_a\partial_j\phi_b ,
\nonumber
\end{alignat}
for $i,j,a,b=1,2$, where $\phi'_3(z)=\partial_z\phi_3(z)$.
On the other hand, on the $(r,\theta)$ surface in the limit $z\rightarrow\pm\infty$,
we have $J^j\intd S_j=J^3\intd x^1\intd x^2$, where 
\begin{alignat}1
J^3=\frac{1}{2\pi}\frac{\pm m_0}{(\Delta^2(r)+m_0^2)^{3/2}}\epsilon^{ab}
\partial_1\phi_a\partial_2\phi_b,
\nonumber
\end{alignat}
for $a,b=1,2$.
Substituting these into Eq. (\ref{IndThe}), we have
\begin{widetext}
\begin{alignat}1
{\rm ind}\,{\cal H}_0=&-\frac{1}{4\pi}\oint \intd x^i\epsilon^{ab}\phi_a\partial_i\phi_b
\int_{-\infty}^{\infty} \intd z\frac{\phi_3'(z)}{(\Delta_0^2+\phi_3^2(z))^{3/2}}
-\frac{2}{4\pi}\int \intd^2 x\frac{m_0}{(\Delta^2(r)+m_0^2)^{3/2}}
\epsilon^{ab}\partial_1\phi_a\partial_2\phi_b,
\label{HalIndCon}
\end{alignat}
\end{widetext}
where $i,a,b=1,2$ and $\intd^2x=\intd x^1\intd x^2$.
The factor 2 in the latter term is due to contribution from the two limits, 
$z\rightarrow\pm\infty$.
The integration over $z$ in the former term can be carried out as follows:
\begin{alignat}1
\int_{-\infty}^{\infty} \intd z\frac{\phi'(z)}{(\Delta_0^2+\phi^2(z))^{3/2}}
&=\int_{-m_0}^{m_0}\intd x(x^2+\Delta_0^2)^{-3/2}
\nonumber\\
&=\frac{2m_0}{\Delta_0^2\sqrt{m_0^2+\Delta_0^2}} .
\nonumber
\end{alignat}
Therefore, the first term becomes, after the integration over $z$,
\begin{alignat}1
-\frac{m_0}{\sqrt{m_0^2+\Delta_0^2}}\frac{1}{2\pi}
\oint \intd x^i\epsilon^{ab}\hat\phi_a\partial_i\hat\phi_b ,
\label{HalForInt}
\end{alignat}
where $\hat{\bm\phi}={\bm\phi}/\Delta_0$ at $r=\infty$.
The second term in (\ref{HalIndCon}) can be calculated as
\begin{alignat}1
&
\frac{1}{2\pi}\oint\intd\theta\int_0^\infty \intd r\frac{m_0}{(\Delta^2(r)+m_0^2)^{3/2}}
\epsilon^{ab}\partial_r\phi_a\partial_\theta\phi_b
\nonumber\\
&
=\frac{1}{2\pi}\oint\intd\theta\epsilon^{ab}\hat\phi_a\partial_\theta\hat\phi_b
\int_0^\infty \intd r\frac{m_0\Delta\Delta'}{(\Delta^2(r)+m_0^2)^{3/2}},
\nonumber
\end{alignat}
where we have used the relation $\phi_a(r,\theta)=\hat\phi_a(\theta)\Delta(r)$, 
which is valid from Eq. (\ref{SCOrdParAsy}).
The integral over $r$ thus reduces to
\begin{alignat}1
\int_0^\infty \intd r\frac{m_0\Delta\Delta'}{(\Delta^2(r)+m_0^2)^{3/2}}
&=
\int_0^{\Delta_0^2} \intd x\frac{m_0}{2(x+m_0^2)^{3/2}}
\nonumber\\
&=\frac{m_0}{\sqrt{m_0^2}}-\frac{m_0}{\sqrt{\Delta_0^2+m_0^2}} .
\nonumber
\end{alignat}
Therefore, the second term in Eq. (\ref{HalIndCon}) becomes
\begin{alignat}1
-\left({\rm sgn}(m_0)-\frac{m_0}{\sqrt{\Delta_0^2+m_0^2}}\right)
\frac{1}{2\pi}\oint \intd\theta\epsilon^{ab}\hat\phi_a\partial_\theta\hat\phi_b .
\label{HalLatInt}
\end{alignat}
From Eqs. (\ref{HalForInt}) and (\ref{HalLatInt}), it turns out that 
Eq. (\ref{HalIndCon}) finally becomes
\begin{alignat}1
{\rm ind}\,{\cal H}_0&=
-{\rm sgn}(m_0)
\frac{1}{2\pi}\oint\intd\theta\epsilon^{ab}\hat\phi_a\partial_\theta\hat\phi_b .
\nonumber\\
&=-{\rm sgn}(m_0)q .
\label{FinIndThe}
\end{alignat}
The right hand side of Eq. (\ref{FinIndThe}), the topological index, is indeed
the same as the analytical index, the left hand side, 
summarized in Table \ref{t:NorWavFun}.
We also note that the first line of Eq. (\ref{FinIndThe}) tells that the present 
index is a product of the sign of $m_0$ and 
the winding number around a vortex which is a topological invariant of a line defect.

\section{Summary and discussions}\label{s:Sum} 

We have studied a 3D model of a topological superconductor with a vortex line.
In addition to particle-hole symmetry which is the most basic symmetry of 
the Bogoliubov-de Gennes Hamiltonian, the model has chiral symmetry
when the chemical potential is zero.
This enables us to explore the Majorana zero modes in terms of the index theorem.

In the former part of the paper, we have solved the Bogoliubov-de Gennes equation
at the zero energy 
for the Hamiltonian with a vorticity-$q$ vortex.
When the chemical potential is zero, 
we have obtained $|q|$ Majorana zero modes. 

To solve the Bogoliubov-de Gennes equation, we have made the best use of the
conserved angular momentum around the $z$ axis. 
Precisely speaking, it may not be suitable to call the zero modes thus obtained 
Majorana modes, since it is not ``real''. To see this, note that
Eq. (\ref{PHTraUm}) allows us to choose $\pm C\Psi_m^\pm= \Psi_{-m}^\pm$.
Namely, ``complex conjugate'' of $\Psi_m^\pm$ is not itself but is $\Psi_{-m}^\pm$ which is 
orthogonal to $\Psi^\pm_m$.
In this case, linear combinations $(\Psi_m^++\Psi_{-m}^+)/\sqrt{2}$ and
${\rm i}(\Psi_m^+-\Psi_{-m}^+)/\sqrt{2}$, and similarly for $\Psi^-_{\pm m}$
can be regarded as Majorana modes. 
Alternatively, one can add rotational symmetry breaking perturbations.
Then, $m$ is not a good quantum number any longer, and one can obtain generically Majorana 
wave functions. Even in such a case, the index theorem guarantees 
that the Majorana zero modes are 
topologically protected and such small perturbations cannot change the index,
which implies at least $|q|$ zero modes even for a perturbed model.

Using $|q|$ zero modes obtained so far and taking their chirality into account,
we have calculated the analytical index.
On the other hand, 
in the case of nonzero chemical potential, we have obtained just one zero mode when $q$ is odd,
whereas no zero modes when $q$ is even.
Therefore, this result suggests that Majorana zero modes are classified by Z for 
systems of class BDI, whereas by Z$_2$ for systems of class D.
\cite{TSL07,GurRad07,CLGS09,FukFuj10}

This classification scheme can also be supported by the following arguments:\cite{PCKane}
Let us consider the case $m_0>0$ and $q>0$, for example. When $\mu=0$,
we have $q$ zero modes whose eigen functions are denoted as $\Psi_m=(0,\Psi_m^-)^T$,
where $|m|\le(q-1)/2$.
For these unperturbed states, we can compute the first order perturbative corrections
of the chemical potential term, or more generically, of some hermitian operator ${\cal O}$
which is odd under particle-hole transformation ${\cal C}{\cal O}{\cal C}^{-1}=-{\cal O}$ 
but even under chiral transformation $\Gamma_5{\cal O}\Gamma_5={\cal O}$. 
Note that we can choose the phase of the states such that ${\cal C}\Psi_m=\Psi_{-m}$.
Therefore, we have
$\langle\Psi_m|{\cal O}|\Psi_m\rangle=-\langle\Psi_{-m}|{\cal O}|\Psi_{-m}\rangle$.
These matrix elements do not vanish in general for operators breaking chiral symmetry
and we have $\pm \varepsilon$ energy 
corrections to these states.\cite{footnote1}
However, there is one exception $m=0$ which occurs when $q$ is odd.
The above equation readily leads to $\langle\Psi_{0}|{\cal O}|\Psi_{0}\rangle=0$.
Therefore, we can expect that $\Psi_0$ state is protected from chiral symmetry breaking
perturbations.

In the latter part of the paper, we have calculated the topological index.
In the previous paper,\cite{FukFuj10} 
we have derived generic index theorem on the implicit assumption
that the defect of the model is point-like. 
However, in 3D superconductors, a vortex is a line defect. 
It has been shown that the index theorem obtained in the previous paper
is valid even in the present case of a line defect. 

It turns out that the index theorem is a quite powerful tool to explore the Majorana zero modes 
bound to topological defects in superconductors. 
However, it needs chiral symmetry by definition. 
Therefore, it may be interesting to extend the index theorem such that
it can be applicable to more generic universality classes of superconductors.

\begin{acknowledgments}
The author would like to thank C. L. Kane for fruitful discussions, especially
on the conserved angular momentum, and 
T. Fujiwara for extensive discussions.
This work was supported in part by Grants-in-Aid for Scientific Research
(Nos. 20340098 and 21540378).
\end{acknowledgments}


\begin{thebibliography}{99}
\bibitem{footnote1}
If ${\cal O}$ is odd under $\Gamma_5$, i.e., $\Gamma_5{\cal O}\Gamma_5=-{\cal O}$, then
$\langle\Psi_m|{\cal O}|\Psi_m\rangle=\langle\Psi_m|\Gamma_5{\cal O}\Gamma_5|\Psi_m\rangle=
-\langle\Psi_m|{\cal O}|\Psi_m\rangle$, where we have used 
$\Gamma_5|\Psi_m\rangle=-|\Psi_m\rangle$.
This leads to 
$\langle\Psi_m|{\cal O}|\Psi_m\rangle=0$.
However, there are generically no reasons for even operators under $\Gamma_5$
to vanish the matrix elements.
\end{thebibliography}
\end{document}